\documentclass[twocolumn,english,prl,amsmath,amsfonts,twocolumn]{revtex4}
\usepackage[T1]{fontenc}
\usepackage[utf8]{inputenc}
\usepackage{amsmath}
\usepackage{graphicx}

\makeatletter

\@ifundefined{textcolor}{}
{%
 \definecolor{BLACK}{gray}{0}
 \definecolor{WHITE}{gray}{1}
 \definecolor{RED}{rgb}{1,0,0}
 \definecolor{GREEN}{rgb}{0,1,0}
 \definecolor{BLUE}{rgb}{0,0,1}
 \definecolor{CYAN}{cmyk}{1,0,0,0}
 \definecolor{MAGENTA}{cmyk}{0,1,0,0}
 \definecolor{YELLOW}{cmyk}{0,0,1,0}
 }

\makeatletter

\usepackage{babel}

\begin{document}

\title{Bloch oscillations of Path-Entangled Photons}

\author{Yaron Bromberg}
\email{yaron.bromberg@weizmann.ac.il}

\author{Yoav Lahini}
\author{Yaron Silberberg}

\affiliation{Department of Physics of Complex Systems, Weizmann Institute of Science,
Rehovot 76100, Israel}

\begin{abstract}

We show that when photons in $N$-particle path entangled $\left|N,0\right\rangle +\left|0,N\right\rangle $
state undergo Bloch oscillations, they exhibit a periodic transition
between spatially bunched and antibunched states. The transition occurs even when the photons
are well separated in space. We study the scaling of the bunching-antibunching period, and show it is proportional to $1/N$.
\end{abstract}

\maketitle

When electrons in crystalline potentials are subjected to uniform
external fields, classical mechanics predicts that they will exhibit Ohmic transport. Remarkably, in 1929 Bloch predicted that the quantum
coherence properties of the electrons prevent their transport \cite{Bloch1928,Ashcroft}.
He showed that the electrons dynamically localize and undergo periodic
oscillations in space. Bloch oscillations (BOs) manifest the wave
properties of the electrons, and therefore appear in other systems
of waves in tilted periodic potentials. BOs were observed for electronic
wavepackets in semiconductor supperlattices \cite{SuperLattice},
matter waves in optical lattices \cite{MatterWaves1} and light waves
in tilted waveguide lattices \cite{Lederer,Roberto}and
in periodic dielectric systems\cite{Wiersma}.

In optics, BOs manifest the classical wave properties of light, and
not its quantum (particle) nature. Recently, quantum properties of
light propagating in periodic lattices of identical waveguides have
been studied, predicting the emergence of nontrivial photon correlations
\cite{Rai08,QWG}. Nonclassical correlations between photon pairs
were experimentally observed in periodic lattices \cite{Peruzzo},
while the effect of disorder was studied in \cite{lahini}. BOs of
a single photon in tilted lattices were shown to follow the dynamics
of coherent states \cite{Rai09Bloch}. Nonclassical features of BOs
of photons in a two-band model were studied by Longhi, who showed
that the probability to detect photon pairs in different bands oscillates
nonclassically \cite{LonghiBloch}.

In this paper we study theoretically the propagation of spatially
entangled states in waveguide lattices which exhibit Bloch oscillations.
We consider light fields initiated in a superposition of $N$ photons
in site \textbf{$\mu'$} or in site $\nu'$, $\left|\psi\right\rangle =\frac{1}{\sqrt{2}}\left(\left|N\right\rangle _{\mu'}\left|0\right\rangle _{\nu'}+e^{-i\varphi}\left|0\right\rangle _{\mu'}\left|N\right\rangle _{\nu'}\right)$.
Such superpositions, coined NOON states, exhibit fascinating quantum
interference properties. NOON states are considered the optimal quantum
states of light for quantum meteorology applications such as quantum
lithography and quantum imaging \cite{Dowling}. Here we show that
when NOON states undergo BOs, the nature of the correlations between
the photons oscillate between spatially bunched and antibunched states.
We find that the period of the oscillations is inversely proportional
to the photon number $N$, resembling the $\lambda/N$ oscillations
of NOON states in Mach-Zehnder interferometers. Interestingly, the
oscillation period is also inversely proportional to the initial separation
of the two input sites $\mu'-\nu'$. A unique feature of the NOON
state BOs is that the transition between the bunched and antibunched
states can happen even when the photons are separated by many lattice
sites.

We consider the simplest waveguide structure which exhibits BOs, a
one-dimensional lattice of single mode waveguides which are evanescently
coupled. The propagation in the lattice is determined by two parameters:
the phase accumulation rate in the waveguides (the propagation constant)
and the tunneling rate between neighboring sites (the coupling constant)
\cite{Hagai98}. The propagation of the fields in waveguide lattices
is described by the tight-binding model, and was used to demonstrate
many optical analogues of solid-state phenomena \cite{ReviewRep,LonghiReview}.
BOs are observed when the coupling constants between all the waveguides
are identical and the propagation constants depend linearly on the
waveguide position \cite{Roberto,Lederer}. To study the propagation
of nonclassical light in such a structures we quantize the fields
in the lattice. Since each of the waveguides supports a single mode,
the field in waveguide $\mu$ is represented by the bosonic creation
and annihilation operators $a_{\mu}^{\dagger}$ and $a_{\mu}$, which
satisfy the commutation relations $[a_{\mu},a_{\nu}^{\dagger}]=\delta_{\mu,\nu}$.
The operators evolve according to the Heisenberg equations \cite{QWG}:

\begin{equation}
-i\frac{\partial a_{\mu}^{\dagger}}{\partial z}=\mu Ba_{\mu}^{\dagger}+C\left(a_{\mu+1}^{\dagger}+a_{\mu-1}^{\dagger}\right).\label{eq:1}\end{equation}

Here $z$ is the spatial coordinate along the propagation axis, $C$
is the coupling constant and $B$ is the difference in the propagation
constants of neighboring sites. The evolution of the creation and
annihilation operators is calculated using the Green function $U_{\mu,\mu'}(z)$
of Eq. (\ref{eq:1}), $a_{\mu}^{\dagger}(z)=\sum_{\mu'}U_{\mu,\mu'}(z)a_{\mu'}^{\dagger}(z=0)$
\cite{QWG}. The unitary transformation $U_{\mu,\mu'}(z)$ describes
the amplitude for the transition of a single photon from waveguide
$\mu$ to waveguide $\mu'$. The Green function of Eq. (\ref{eq:1})
is given by \cite{Lederer,NJP04}:\begin{equation}
U_{\mu,\mu'}(z)=e^{i\frac{\pi}{2}(\mu'-\mu)}e^{i\frac{Bz}{2}(\mu'+\mu)}J_{\mu'-\mu}\left(\tfrac{4C}{B}sin\left(Bz/2\right)\right),\label{eq:2}\end{equation}
 where $J_{\mu}(x)$ is the $\mu$th Bessel function of the first
kind. Since any input state can be expressed with the creation operators
$a_{\mu}^{\dagger}$ and the vacuum state $\left|0\right\rangle $,
the evolution of nonclassical states along the lattice can be calculated
using Eq. (\ref{eq:2}). The probability to locate at site $\mu$
a photon that is injected into the lattice at site $\mu'=0$ is given
by the photon density $n_{\mu}=\left\langle a_{\mu}^{\dagger}a_{\mu}\right\rangle =|U_{\mu,\mu'=0}|^{2}$
and is depicted in Fig. 1(a). The photon exhibits BOs: it spreads
across the lattice by coupling from one waveguide to its neighbors
in a pattern characterized by two peaks at the two edges of the distribution. Each peak covers approximately four waveguides, and and its location oscillates along the propagation axis with a period $\lambda_{B}=2\pi/B$.
The right and left paths of the two peaks mark the two branches of the BO. We note that such
a double-branch pattern is not a special feature of single photons.
Any state of light which is coupled to a single waveguide in the lattice
will exhibit exactly the same photon density. However, when the light
is coupled to more than one waveguide, the propagation of the photons
becomes state-dependent. Rai et al. have shown that a single photon
which is coupled into the lattice in a superposition of several waveguides
exhibits BOs like a coherent state \cite{Rai09Bloch}. Figure 1(b)
shows the photon density for a superposition of a single photon initiated
in a superposition of two neighboring waveguides, with a relative
phase $\varphi=0$. The two paths the photon can take, starting either
from waveguide $\mu'=0$ or from waveguide $\nu'=1$, contribute coherently
to the photon density $n_{\mu}=\tfrac{1}{2}|U_{\mu,\mu'=0}+U_{\mu,\nu'=1}|^{2}$.
Because of this interference the photon oscillates in a single branch,
exactly like a coherent beam. In contrast, when a NOON state with
$N>1$ is coupled to the lattice, the photon density is identical
to the photon density obtained by two incoherent beams $n_{\mu}=\tfrac{N}{2}|U_{\mu,\mu'=0}(z)|^{2}+\tfrac{N}{2}|U_{\mu,\nu'=1}(z)|^{2}$
(Fig. 1(c)).

\begin{figure}
\includegraphics[clip,width=1\columnwidth]{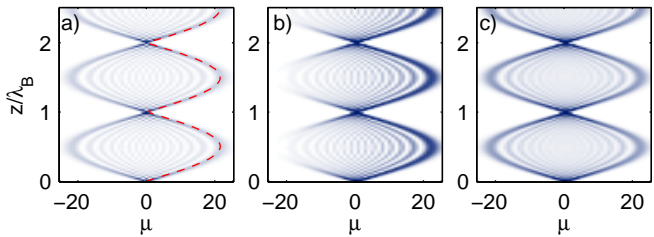}

\caption{(Color online) (a) The photon density $n_{\mu}(z)=\left\langle a_{\mu}^{\dagger}a_{\mu}\right\rangle $
for a single photon initiated at the waveguide $\mu'=0$. The photon
is mostly localized in two narrow peaks at the edges of the distribution.
The path of each peak has a period $\lambda_{B}$, and marks a branch
of the Bloch oscillation (red dashed line marks the right branch).
(b) The photon density $n_{\mu}(z)$ for a $N00N$ state input with
$N=1$ coupled to waveguides $\mu'=0$ and $\nu'=1$. The photon is
mainly located at the right branch of the oscillation. (c) The photon
density $n_{\mu}(z)$ for a $N00N$ input state with $N=2$ coupled
to waveguides $\mu'=0$ and $\nu'=1$ The photons from $\mu'=0$ and
$\nu'=1$ inputs add up incoherently, showing double-branch oscillations.}

\end{figure}

\begin{figure}
\includegraphics[clip,width=1\columnwidth]{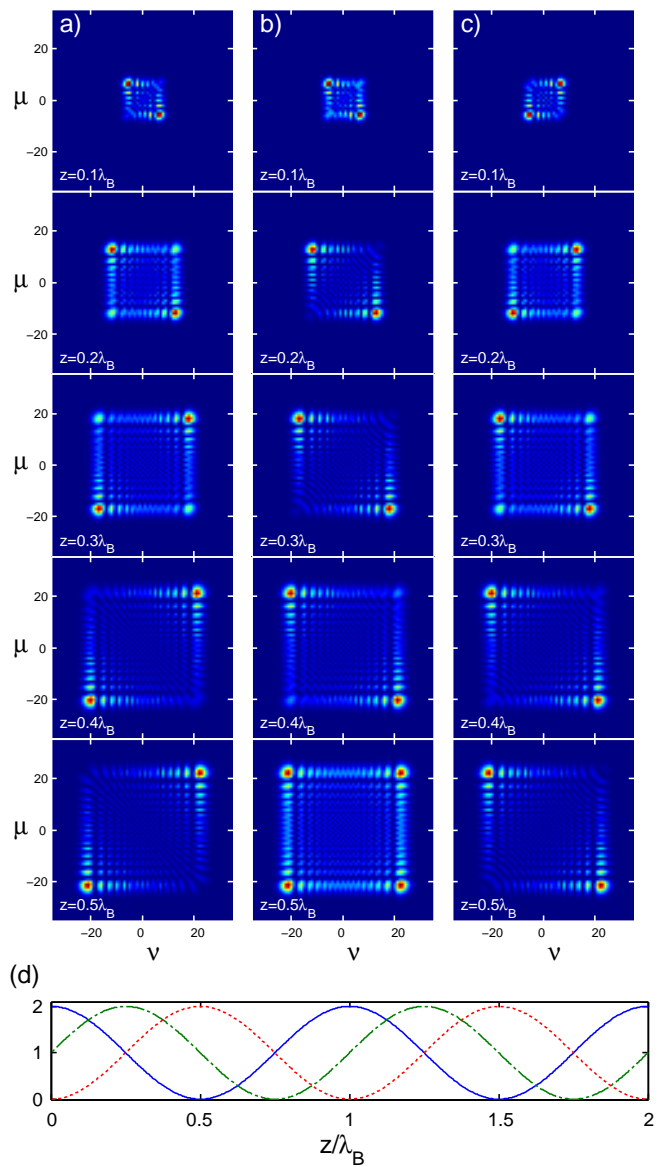}\label{fig:2}

\caption{(Color online) Bloch oscillations of NOON states with $N=2$ coupled
to two adjacent waveguides $\left|\psi\right\rangle =\tfrac{1}{\sqrt{2}}(\left|2\right\rangle _{0}\left|0\right\rangle _{1}+e^{-i\varphi}\left|0\right\rangle _{0}\left|2\right\rangle _{1})$.
(a) The multiple detection probability $\Gamma_{\mu,\nu}^{(1,1)}$
at several propagation distances, for $\varphi=0$. At the beginning
of the propagation the two photons exhibit antibunching and are located
at the two different branches of the oscillations. As the photons
approach the turning point $(z=\lambda_{B}/2),$ they bunch and are
found with the highest probability in the same branch. (b) Same as
(a) for $\varphi=\frac{\pi}{2}$. The photons show bunching-antibunching
cycle, but in this case start the oscillation partially bunched. (c)
Same as (a) and (b), for $\varphi=\pi$. Here the photons start the
bunching-antibunching cycle bunched. (d) The normalized coincidence
rate $\gamma^{(1,1)}(z)$ as a function of the lattice length for
$\varphi=0$ (blue solid line), $\varphi=\tfrac{\pi}{2}$ (green dash-dotted
line), and $\varphi=\pi$ (red dottde line). The coincidence rate
is calculated between the positions of the central waveguide in each
branch, showing oscillations with a period $\lambda_{B}$. }

\end{figure}

Nonclassical features of light are probed by correlations between
the photons. We focus on the probability to detect $p$ photons in
waveguide $\mu$ and $q=N-p$ photons in waveguide $\nu$, $\Gamma_{\mu,\nu}^{(p,q)}=\tfrac{1}{q!p!}\left\langle a_{\mu}^{\dagger^{p}}a_{\nu}^{\dagger^{q}}a_{\nu}^{q}a_{\mu}^{p}\right\rangle $
\cite{MandelWolf}. For a NOON state coupled to waveguides $\mu'$and
$\nu'$, the multiple detection probability is:

\begin{eqnarray}
\Gamma_{\mu,\nu}^{(p,q)} & = & \left.\tfrac{1}{2}\tfrac{N!}{p!(N-p)!}\right|J_{\mu'-\mu}(\zeta)^{p}J_{\mu'-\nu}(\zeta)^{q}\label{eq:3}\\
 & + & \left.e^{i\theta(z)}J_{\nu'-\mu}(\zeta)^{p}J_{\nu'-\nu}(\zeta)^{q}\right|^{2}\nonumber \end{eqnarray}

Where $\zeta=4C/B\, sin\left(Bz/2\right)$, and $\theta(z)$ is given
by:\begin{equation}
\theta(z)=\varphi+\frac{1}{2}(\pi+Bz)(\nu'-\mu')N\:.\label{eq:4}\end{equation}

\begin{figure}
\label{fig:3}\includegraphics[clip,width=1\columnwidth]{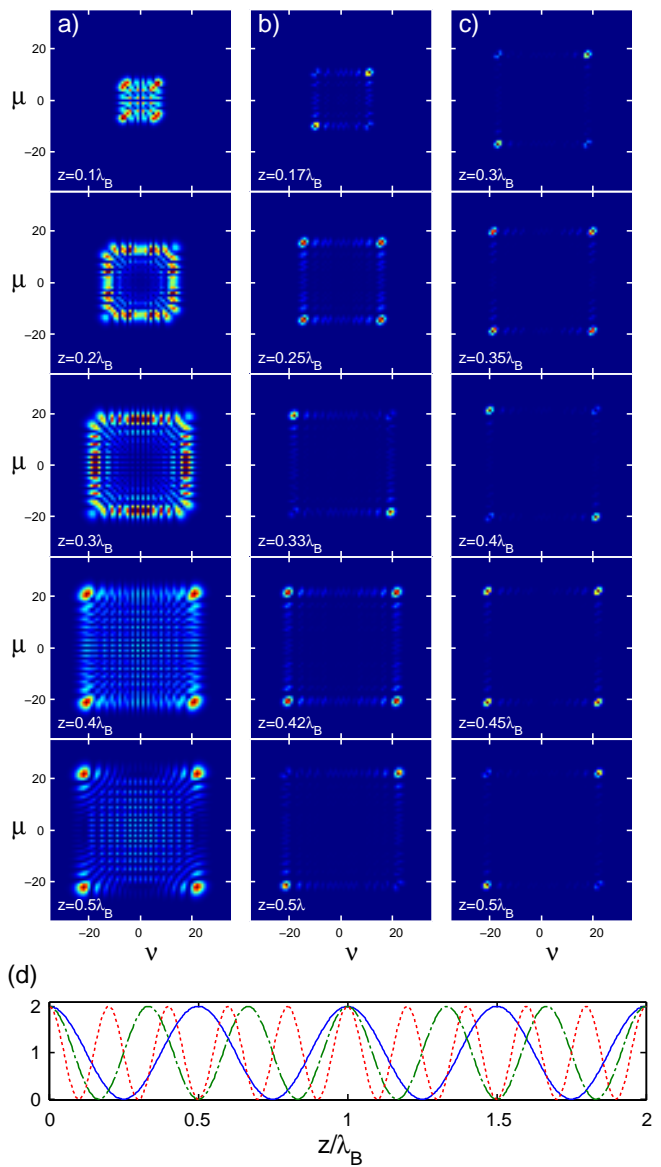}

\caption{(Color online) Bloch oscillations of NOON states with sub-$\lambda_{B}$
correlation-oscillation periods. (a) The multiple detection probability
$\Gamma_{\mu,\nu}^{(1,1)}$ at several propagation distances for the
input state $\left|\psi\right\rangle =\tfrac{1}{\sqrt{2}}(\left|2\right\rangle _{-1}\left|0\right\rangle _{1}+\left|0\right\rangle _{-1}\left|2\right\rangle _{1})$.
The photons exhibit bunching-antibunching oscillations (see text)
with a period $\lambda_{B}/2.$ (b),(c) The multiple detection probability
$\Gamma_{\mu,\nu}^{(N/2,N/2)}$ for a NOON state with $N=6$ (b) and
$N=10$ (c), injected to adjacent waveguides $\left|\psi\right\rangle =\tfrac{1}{\sqrt{2}}(\left|N\right\rangle _{0}\left|0\right\rangle _{1}+\left|0\right\rangle _{0}\left|N\right\rangle _{1})$.
The oscillations of the correlation matrix are much faster, hence
the probability matrix is calculated for five lattice lengths close
to the turning point $z=\lambda_{B}/2$. 
(d) The normalized coincidence rate $\gamma^{(N/2,N/2)}(z)$ as a
function of the lattice length for the above three cases. The period
of the oscillations are $\lambda_{B}/2$ {[}(a), blue solid line{]}
$\lambda_{B}/3$ {[}(b), green dashed-dotted{]} and $\lambda_{B}/5$
{[}(c), red dotted line{]}. }
\end{figure}
Eq. (\ref{eq:3}) shows that two terms contribute to the multiple
detection probability: the photons arrive either from the input waveguide
$\mu'$ or from waveguide $\nu'$. Since the photons are indistinguishable,
these two paths interfere. The phase between the two paths is proportional
to $Nz$, indicating that the oscillation period scales like $1/N$
(see below).

In Fig. 2 we depict $\Gamma_{\mu,\nu}^{(1,1)}$, the probability to
detect one photon at waveguide $\mu$ and another photon at waveguide
$\nu$, for a NOON state with $N=2$. The left column shows the evolution
of the probability for a NOON state with a phase $\varphi=0$. At
the beginning of the propagation (\textbf{$Bz\ll\pi$}), the photons
follow the same path as in a periodic array of identical waveguides
\cite{QWG,Peruzzo}. At this stage the off-diagonal terms of the probability
matrix $\Gamma_{\mu,\nu}^{(1,1)}$are much stronger than the diagonal
terms, indicating that the photons exhibit antibunching: each photon
takes a different branch of the oscillation. However, during the expansion
period of the BO, as the photons approach the turning point, the symmetry
of the two-photon probability matrix changes significantly. The diagonal
terms of the matrix become more pronounced, i.e. there is a higher
probability to find the two photons in the same branch.
At the turning point $z=\tfrac{\lambda_{B}}{2}$, the photons bunch:
the off-diagonal terms of the probability matrix vanish, indicating
that the photons are never found simultaneously at the two different
branches. Remarkably, even though the photons start the propagation
in spatially separated branches, at the turning point they bunch to
one of the branches, with equal probability. Beyond this point the
photon density contracts back towards the origin waveguides. During
this contraction the pairs again switch to an antibunched state. We
note that the bunching-antibunching transition happens when the two
branches of the BO are spatially separated, whereas the bunching-antibunching
transition predicted in binary lattices occurs only when the photons
are in the same waveguide \cite{LonghiBloch}. The cycle in the symmetry
of the probability matrix is observed for any initial phase of the
NOON state phase $\varphi$, as demonstrated Fig. 2(b) and 2(c). The
phase $\varphi$ sets how bunched or antibunched the photons are at
the beginning of the propagation, but the period of the cycle is phase-independent
{[}see Eq. (\ref{eq:4}){]}.

The bunching-antibunching cycle described above can be realized experimentally
by measuring the correlations at the output of lattices with identical
parameters but with different propagation lengths. For each propagation
length, the waveguide at the center of each oscillating branch (henceforth
waveguides $x$ and $y$) can be imaged on two photon-number resolving
detectors. The probability to detect $p$ photons at waveguide $x$
and $q=N-p$ photons at waveguide $y$ is proportional $\Gamma_{x,y}^{(p,q)}$.
When a delay is introduced between the photons that are injected to
waveguide $\mu'$ and the photons injected to waveguide $\nu'$, the
photons become distinguishable, as in the Hong-Ou-Mandel (HOM) experiment
\cite{HOM}. This corresponds to replacing the NOON state with a mixed
state of $N$ photons in either one of the two input waveguides. The
ratio of the detection probabilities for the N00N and mixed states
is given by

\begin{equation}
\gamma^{(p,q)}=\frac{\Gamma_{x,y}^{(p,q)}}{\frac{1}{2}\frac{N!}{p!(N-p)!}\left(|J_{\mu'-x}^{p}J_{\mu'-y}^{q}|^{2}+|J_{\nu'-x}^{p}J_{\nu'-y}^{q}|^{2}\right)}.\label{eq:5}\end{equation}

Figure 2(d) shows $\gamma^{(1,1)}$ as a function of the lattice length,
for NOON states with $N=2$ and $\varphi=0,\pi/2,\,\pi$. When $\gamma^{(1,1)}=0$,
the photons are bunched and are never found in the two different branches
of the BO; scanning the delay between the input ports of the lattice
will yield a HOM dip. When $\gamma^{(1,1)}=2$, the photons are antibunched,
and a delay scan will result in a HOM peak \cite{HOMShaping}. Figure
2(d) clearly shows that the bunching-antibunching oscillations have
a period of $\lambda_{B}$.

We next study input states which exhibit correlation oscillations
with shorter periods. Eq. (\ref{eq:4}) suggests that the period of
the oscillations in the correlation properties depends on the spacing
between the input waveguides and on the number of photons in the NOON
state. Figure 3 shows several examples of correlation oscillations
with periods shorter than $\lambda_{B}$. In Fig. 3a we show the propagation
for a NOON state with $N=2$, where the input sites are separated
by one waveguide. In this case the photons exhibit a bunching-antibunching
transition with a different spatial symmetry \cite{QWG}. The correlation
map oscillates between a state in which the peaks are highest at the
corners of the correlations matrix, to a case in which the highest
probability is between the corners. The oscillation period is $\lambda_{B}/2$,
twice the period observed for a NOON state input with adjacent waveguides.
Finally we calculate $\Gamma_{\mu,\nu}^{(N/2,N/2)}$ for NOON states
with $N=6$ {[}Fig. 3(b){]} and $N=10$ {[}Fig. 3(c){]}, with adjacent
input waveguides. The oscillation period is indeed $2\lambda_{B}/N$,
as predicted by Eq. (\ref{eq:4}). Within one oscillation of the single
photon density, the $N$-photon distribution switches $N/2$ times
from all the photons in the same branch to photons divided equally
between the two branches.

In conclusion, we studied the propagation of photonic NOON states
in waveguide lattices which exhibit Bloch oscillations. We found that
while the photon density oscillates in the Bloch frequency, the multiple
detection probability oscillates at higher frequencies. These oscillations
indicate that the photons show a transition from a bunched to antibunched
states, with a period that scales as $1/N$. By carefully designing
the parameters of the Bloch lattice this oscillatory transition can
be used to distribute bunched and antibunched states of light in an
integrated and thus robust manner. To experimentally observe the bunching-antibunching
transition, we propose to perform a Hong-Ou-Mandel measurement between
two waveguides at the two branches of oscillations using photon-number
resolving detectors. We predict oscillations between a HOM dip and
peak as a function of the propagation distance in the lattice. Recent
progress in waveguide lattice fabrication \cite{Obrien,SzameitWGs},
photon number resolving detectors and photonic NOON state sources
\cite{ItaiMPPC}, make such measurements in reach.

We thank Shlomi Kotler for fruitful discussions, and Gang Wang from the Chinese University of
Hong Kong for bringing our attention to an error in the original manuscript. Financial support
by the Minerva Foundation and the Crown Photonics Center is gratefully
acknowledged. Y. L. is supported by the Israeli Academy of Science
and Humanities through the Adams program.


\begin{thebibliography}{24}
\bibitem{Bloch1928}F. Bloch, Z. Phys. \textbf{52}, 555 (1928).

\bibitem{Ashcroft}N.W. Ashcroft and N.D. Mermin, Solid State Physics,
(Holt-Saunders Int. Ed., Philadelphia, 1981).

\bibitem{SuperLattice}J. Feldmann et al., Phys. Rev. B 46, R7252
(1992); K. Leo et al., Solid State Commun. \textbf{84}, 943 (1992).

\bibitem{MatterWaves1} M. BenDahanet al., Phys. Rev. Lett. \textbf{76},
4508 (1996);S. R. Wilkinson et al., Phys. Rev. Lett. \textbf{76},
4512 (1996) .

\bibitem{Roberto}R. Morandotti et al., Phys. Rev. Lett. \textbf{83},
4756 (1999).

\bibitem{Lederer}T. Pertsch et al., Phys. Rev. Lett. \textbf{83},
4752 (1999).

\bibitem{Wiersma} R. Sapienza et al., Phys. Rev. Lett. \textbf{91},
263902 (2003).

\bibitem{Rai08}A. A. Rai, G. S. Agarwal, and J. H. H. Perk, Phys.
Rev. A \textbf{78,} 042304 (2008).

\bibitem{QWG}Y. Bromberg, et al., Phys. Rev. Lett. \textbf{102},
253904 (2009).

\bibitem{Peruzzo}A. Peruzzo et al., Science \textbf{329}, 1500 (2010)

\bibitem{lahini}Y. Lahini, et al., Phys. Rev. Lett. \textbf{105},
163905 (2010).

\bibitem{Rai09Bloch}A.Rai and G. S. Agarwal Phys. Rev. A \textbf{79},
053849 (2009).

\bibitem{LonghiBloch}S. Longhi, Phys. Rev. Lett. \textbf{101}, 193902
(2008); S. Longhi, Phys. Rev. B \textbf{79}, 245108 (2009).

\bibitem{Dowling}J. Dowling, Contemporary Physics 49, 125 (2008).

\bibitem{Hagai98}H. S. Eisenberg et al., Phys. Rev. Lett. \textbf{81},
3383 (1998).

\bibitem{ReviewRep} F. Lederer et al., Phys. Rep. \textbf{463}, 1-126
(2008).

\bibitem{LonghiReview}S. Longhi, Laser \& Photon. Rev. \textbf{3},
243–261 (2009).

\bibitem{NJP04}T. Hartmann, et al., New J. Phys. \textbf{6, }2 (2004).

\bibitem{MandelWolf}L. Mandel and E. Wolf, Optical Coherence and
Quantum Optics (Cambridge, United Kingdom, 1995).

\bibitem{HOM}C. K. Hong, Z. Y. Ou, and L. Mandel, Phys. Rev. Lett.
\textbf{59}, 2044(1987).

\bibitem{HOMShaping}B. Dayan, Y. Bromberg, I. Afek, and Y. Silberberg,
Phys. Rev. A 75, 043804 (2007).

\bibitem{Obrien}A. Politi et al., Science \textbf{320,} 646 – 649
(2008). G. D. Marshall et al., Opt. Express \textbf{17}, 546(2009).

\bibitem{SzameitWGs}F. Dreisow et al., Phys. Rev. Lett. \textbf{102},
076802 (2009).

\bibitem{ItaiMPPC}I. Afek, et al., Phys. Rev. A \textbf{79}, 043830
(2009); I. Afek, O. Ambar and Y. Silberberg, Science 328, 879 (2010).

\end{thebibliography}
\end{document}